\documentclass{emulateapj}
\usepackage{amsmath}

\newcommand{\actaa}{Acta Astron.}

\newcommand{\bdv}[1]{\mbox{\boldmath$#1$}}

\def\au{{\rm AU}} 
 
\def\kms{{\rm km}\,{\rm s}^{-1}}
\def\masyr{{\rm mas}\,{\rm yr}^{-1}}
\def\kpc{{\rm kpc}}
\def\mas{{\rm mas}}
\def\muas{{\mu\rm as}}

\def\max{{\rm max}}
\def\rel{{\rm rel}}

\def\hel{{\rm hel}}
\def\geo{{\rm geo}}
\def\e{{\rm E}}
\def\bpi{{\bdv\pi}}
\def\bmu{{\bdv\mu}}

\def\bv{{\bf v}}

\begin{document}
\title{\emph{Spitzer} Microlensing Program as a Probe for Globular Cluster Planets. \\
Analysis of OGLE-2015-BLG-0448}


\author{Rados\l{}aw Poleski\altaffilmark{1,2}, Wei Zhu\altaffilmark{1}, Grant W. Christie\altaffilmark{3}, Andrzej Udalski\altaffilmark{2}, 
Andrew Gould\altaffilmark{1}, Etienne Bachelet\altaffilmark{4}, Jesper Skottfelt\altaffilmark{5}, Sebastiano Calchi Novati\altaffilmark{6,7,8}, 
\\ and\\
M.~K. Szyma\'nski\altaffilmark{2}, I. Soszy\'nski\altaffilmark{2}, G. Pietrzy\'nski\altaffilmark{2,9}, \L{}. Wyrzykowski\altaffilmark{2}, 
K. Ulaczyk\altaffilmark{2,10}, P.~Pietrukowicz\altaffilmark{2}, Szymon~Koz\l{}owski\altaffilmark{2}, J. Skowron\altaffilmark{2}, P. Mr\'oz\altaffilmark{2}, M. Pawlak\altaffilmark{2} 
\\ (OGLE group),\\
C. Beichman\altaffilmark{6}, G. Bryden\altaffilmark{11}, S. Carey\altaffilmark{11}, M. Fausnaugh\altaffilmark{1}, B.~S. Gaudi\altaffilmark{1}, 
C.~B. Henderson\altaffilmark{11,A}, R.~W. Pogge\altaffilmark{1}, Y.~Shvartzvald\altaffilmark{11,A}, 
B. Wibking\altaffilmark{1}, J.~C.~Yee\altaffilmark{12,B} 
\\ (Spitzer team),\\
T.~G. Beatty\altaffilmark{13}, J.~D. Eastman\altaffilmark{12}, J. Drummond\altaffilmark{14}, M.~Friedmann\altaffilmark{15}, 
M. Henderson\altaffilmark{16}, J.~A.~Johnson\altaffilmark{12}, S. Kaspi\altaffilmark{15}, D. Maoz\altaffilmark{15}, J.~McCormick\altaffilmark{17}, 
N. McCrady\altaffilmark{16}, T. Natusch\altaffilmark{3,18}, H.~Ngan\altaffilmark{3}, I. Porritt\altaffilmark{19}, H.~M.~Relles\altaffilmark{20}, 
D.~H. Sliski\altaffilmark{21}, T.-G. Tan\altaffilmark{22}, R.~A.~Wittenmyer\altaffilmark{23,24,25}, J.~T. Wright\altaffilmark{13} 
\\ ($\mu$FUN group),\\
R. A. Street\altaffilmark{4}, Y. Tsapras\altaffilmark{26}, D.~M.~Bramich\altaffilmark{27},
K.~Horne\altaffilmark{28}, C. Snodgrass\altaffilmark{29}, I.~A. Steele\altaffilmark{30}, J.~Menzies\altaffilmark{31}, R.~Figuera~Jaimes\altaffilmark{28,32}, 
J.~Wambsganss\altaffilmark{26}, R. Schmidt\altaffilmark{26}, A. Cassan\altaffilmark{33}, C.~Ranc\altaffilmark{33}, S.~Mao\altaffilmark{34} 
\\ (RoboNet project),\\
V. Bozza\altaffilmark{7,35}, 
M. Dominik\altaffilmark{28,36},
M.~P.~G. Hundertmark\altaffilmark{5}, 
U.~G. J{\o}rgensen\altaffilmark{5}, 
M.~I. Andersen\altaffilmark{37}, 
M.~J. Burgdorf\altaffilmark{38}, 
S.~Ciceri\altaffilmark{39}, 
G. D'Ago\altaffilmark{7,8,35}, 
D. F. Evans\altaffilmark{40}, 
S.-H. Gu\altaffilmark{41}, 
T. C. Hinse\altaffilmark{42}, 
N.~Kains\altaffilmark{43},  
E. Kerins\altaffilmark{43}, 
H. Korhonen\altaffilmark{44,5}, 
M.~Kuffmeier\altaffilmark{5}, 
L. Mancini\altaffilmark{39}, 
A.~Popovas\altaffilmark{5},  
M.~Rabus\altaffilmark{45}, 
S. Rahvar\altaffilmark{46}, 
R.~T. Rasmussen\altaffilmark{47}, 
G. Scarpetta\altaffilmark{7,8}, 
J.~Southworth\altaffilmark{40}, 
J. Surdej\altaffilmark{48}, 
E.~Unda-Sanzana\altaffilmark{49}, 
P. Verma\altaffilmark{8}, 
C. von Essen\altaffilmark{47}, 
Y.-B.~Wang\altaffilmark{41}, 
O. Wertz\altaffilmark{48}
\\ (MiNDSTEp group)}

\email{poleski.1@osu.edu}
\altaffiltext{1}{Department of Astronomy, Ohio State University, 140 W. 18th Ave., Columbus, OH 43210, USA}
\altaffiltext{2}{Warsaw University Observatory, Al. Ujazdowskie 4, 00-478 Warszawa, Poland}
\altaffiltext{3}{Auckland Observatory, Auckland, New Zealand} 
\altaffiltext{4}{Las Cumbres Observatory Global Telescope Network, 6740 Cortona Drive, suite 102, Goleta, CA 93117, USA}
\altaffiltext{5}{Niels Bohr Institute \& Centre for Star and Planet Formation, University of Copenhagen, {\O}ster Voldgade 5, 1350 - Copenhagen K, Denmark}
\altaffiltext{6}{NASA Exoplanet Science Institute, MS 100-22, California Institute of Technology, Pasadena, CA 91125, USA}
\altaffiltext{7}{Dipartimento di Fisica ``E.R. Caianiello'', Universit\`a di Salerno, Via Giovanni Paolo II 132, 84084, Fisciano, Italy}
\altaffiltext{8}{Istituto Internazionale per gli Alti Studi Scientifici (IIASS), Via G. Pellegrino 19, 84019 Vietri sul Mare (SA), Italy}
\altaffiltext{9}{Universidad de Concepci\'on, Departamento de Astronomia, Casilla 160–C, Concepci\'on, Chile}
\altaffiltext{10}{Department of Physics, University of Warwick, Coventry CV4 7AL, UK}
\altaffiltext{11}{Jet Propulsion Laboratory, California Institute of Technology, 4800 Oak Grove Drive, Pasadena, CA 91109, USA}
\altaffiltext{12}{Harvard-Smithsonian Center for Astrophysics, 60 Garden Street, Cambridge, MA 02138, USA}
\altaffiltext{13}{Department of Astronomy and Astrophysics and Center for Exoplanets and Habitable Worlds, The Pennsylvania State  University, University Park, PA 16802, USA}
\altaffiltext{14}{Possum Observatory, Patutahi, New Zealand}
\altaffiltext{15}{School of Physics and Astronomy, Tel-Aviv University, Tel-Aviv 69978, Israel}
\altaffiltext{16}{Department of Physics and Astronomy, University of Montana, 32 Campus Drive, No.  1080, Missoula, MT 59812, USA}
\altaffiltext{17}{Farm Cove Observatory, Centre for Backyard Astrophysics, Pakuranga, Auckland, New Zealand}
\altaffiltext{18}{AUT University, Auckland, New Zealand}
\altaffiltext{19}{Turitea Observatory, Palmerston North, New Zealand}
\altaffiltext{20}{Citizen Scientist}
\altaffiltext{21}{The University of Pennsylvania, Department of Physics and Astronomy, Philadelphia, PA, 19104, USA}
\altaffiltext{22}{Perth Exoplanet Survey Telescope, Perth, Australia}
\altaffiltext{23}{School of Physics and Australian Centre for Astrobiology, UNSW Australia, Sydney, NSW 2052, Australia}
\altaffiltext{24}{Australian Centre for Astrobiology, University of New South Wales, Sydney 2052, Australia}
\altaffiltext{25}{Computational Engineering and Science Research Centre, University of Southern Queensland, Toowoomba, Queensland 4350, Australia}
\altaffiltext{26}{Astronomisches Rechen-Institut, Zentrum f{\"u}r Astronomie der Universit{\"a}t Heidelberg (ZAH), 69120 Heidelberg, Germany} 
\altaffiltext{27}{Qatar Environment and Energy Research Institute(QEERI), HBKU, Qatar Foundation, Doha, Qatar}
\altaffiltext{28}{SUPA, School of Physics \& Astronomy, University of St Andrews, North Haugh, St Andrews KY16 9SS, UK}
\altaffiltext{29}{Planetary and Space Sciences, Department of Physical Sciences, The Open University, Milton Keynes, MK7 6AA, UK}
\altaffiltext{30}{Astrophysics Research Institute, Liverpool John Moores University, Liverpool CH41 1LD, UK}
\altaffiltext{31}{South African Astronomical Observatory, PO Box 9, Observatory 7935, South Africa}
\altaffiltext{32}{European Southern Observatory, Karl-Schwarzschild-Str. 2, 85748 Garching bei M\"unchen, Germany}
\altaffiltext{33}{Sorbonne Universit\'es, UPMC Univ Paris 6 et CNRS, UMR 7095, Institut d'Astrophysique de Paris, 98 bis bd Arago, 75014 Paris, France}
\altaffiltext{34}{National Astronomical Observatories, Chinese Academy of Sciences, 100012 Beijing, China}
\altaffiltext{35}{Istituto Nazionale di Fisica Nucleare, Sezione di Napoli, Napoli, Italy}
\altaffiltext{36}{Royal Society University Research Fellow}
\altaffiltext{37}{Niels Bohr Institute, University of Copenhagen, Juliane Maries Vej 30, 2100 K{\o}benhavn {\O}, Denmark}
\altaffiltext{38}{Meteorologisches Institut, Universit{\"a}t Hamburg, Bundesstra\ss{}e 55, 20146 Hamburg, Germany}
\altaffiltext{39}{Max Planck Institute for Astronomy, K{\"o}nigstuhl 17, 69117 Heidelberg, Germany}
\altaffiltext{40}{Astrophysics Group, Keele University, Staffordshire, ST5 5BG, UK}
\altaffiltext{41}{Yunnan Observatories, Chinese Academy of Sciences, Kunming 650011, China}
\altaffiltext{42}{Korea Astronomy \& Space Science Institute, 776 Daedukdae-ro, Yuseong-gu, 305-348 Daejeon, Republic of Korea}
\altaffiltext{43}{Jodrell Bank Centre for Astrophysics, School of Physics and Astronomy, University of Manchester, Oxford Road, Manchester M13 9PL, UK}
\altaffiltext{44}{Finnish Centre for Astronomy with ESO (FINCA), V{\"a}is{\"a}l{\"a}ntie 20, FI-21500 Piikki{\"o}, Finland}
\altaffiltext{45}{Instituto de Astrof\'isica, Facultad de F\'isica, Pontificia Universidad Cat\'olica de Chile, Av. Vicu\~na Mackenna 4860, 7820436 Macul, Santiago, Chile}
\altaffiltext{46}{Department of Physics, Sharif University of Technology, PO Box 11155-9161 Tehran, Iran}
\altaffiltext{47}{Stellar Astrophysics Centre, Department of Physics and Astronomy, Aarhus University, Ny Munkegade 120, 8000 Aarhus C, Denmark}
\altaffiltext{48}{Institut d'Astrophysique et de G\'eophysique, All\'ee du 6 Ao\^ut 17, Sart Tilman, B\^at. B5c, 4000 Li\`ege, Belgium}
\altaffiltext{49}{Unidad de Astronom{\'{\i}}a, Fac. de Ciencias B{\'a}sicas, Universidad de Antofagasta, Avda. U. de Antofagasta 02800, Antofagasta, Chile}
\altaffiltext{A}{NPP fellow} 
\altaffiltext{B}{Sagan Fellow}

\begin{abstract}
The  microlensing event OGLE-2015-BLG-0448 was observed by \emph{Spitzer} 
and lay within the tidal radius of the globular cluster NGC 6558. 
The event had moderate magnification and was intensively observed, 
hence it had the potential to probe the distribution of planets in globular clusters. 
We measure the proper motion of NGC 6558 
($\bmu_{cl}(N,E) = (+0.36\pm 0.10,+1.42\pm 0.10)\,\masyr$) 
as well as the source and show that the lens is not a cluster member. 
Even though this particular event does not probe the distribution of planets in globular clusters, 
other potential cluster lens events can be verified using our methodology. 
Additionally, we find that microlens parallax measured using OGLE photometry 
is consistent with 
the value found based on the light curve displacement between Earth and \emph{Spitzer}. 
\end{abstract}


\section{{Introduction}
\label{sec:intro}}

\begin{figure}
\plotone{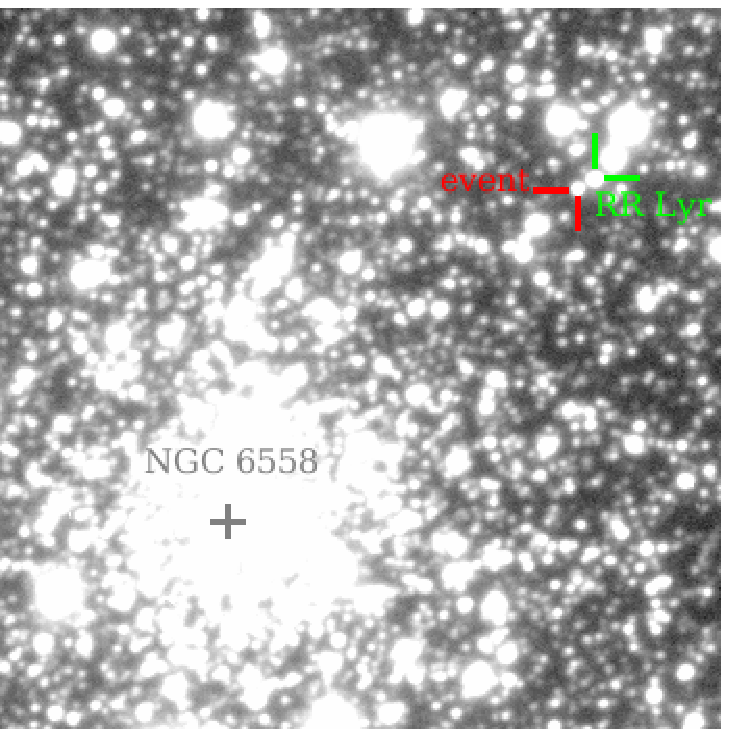}
\caption{Finding chart for OGLE-2015-BLG-0448. 
We marked the center of the globular cluster NGC 6558 (core radius and half-light radius of $0\farcm03$ and $2\farcm15$, respectively), 
the event (baseline brightness $I=16.34~{\rm mag}$), 
and the neighboring RR Lyr star OGLE-BLG-RRLYR-14873 (mean brightness $I=15.52~{\rm mag}$). 
North is up, East is left. The image is $1\farcm5\times1\farcm5$.
}
\label{fig:find}
\end{figure}

The \emph{Spitzer} gravitational microlensing project has as its principal aim
the determination of the Galactic distribution of planets 
\citep{spitzer2015prop}.  This primarily means using \emph{Spitzer}
to measure ``microlens parallaxes'' $\bpi_\e$ and thereby estimate
the distances of the individual lenses. By comparing this overall 
distance distribution to the one restricted to events showing
planetary signatures one can determine whether planets are more
common in, for example, the Galactic disk or the bulge \citep{21event,yee15}. 
Among the 170 microlensing events observed during the 2015 campaign \citep{calchinovati15b},
one event showed a potential for a very different probe of the ``Galactic
distribution of planets'', namely of the frequency of planets in globular
clusters (relative to disk or bulge stars).  The event OGLE-2015-BLG-0448
lay projected against the globular cluster NGC 6558 (Fig.~\ref{fig:find}), and therefore 
the lens was potentially a member of this cluster. 
The lens mass is measured if one knows the relative lens-source parallax and the angular size of the Einstein ring radius \citep{refsdal64}. 
In the case of a globular cluster lens, 
one can in principle derive the lens mass based on the Einstein 
timescale measurement alone \citep[knowing the cluster distance 
and proper motion from the literature; ][]{pac94}. 
In reality, significant uncertainties 
are introduced by the dispersion of bulge source proper motions 
that is comparable to the cluster proper motion. 

Here we present a new method to determine whether the lens from a microlensing
event seen projected against a cluster is in fact a cluster member,
employing observations of the \emph{Spitzer} spacecraft as a
``microlensing parallax satellite''. 
The method is to compare the direction of the heliocentric projected
velocity $\tilde\bv_\hel$ with that of the proper motion of the
cluster relative to the microlensed source $\bmu_{cl,s}$.  As is well
known, $\tilde\bv_\hel$ can be subject to a four-fold degeneracy in
direction \citep{refsdal66,gould94}, but within those degenerate solutions
can be very precisely measured by a parallax satellite \citep{21event}.
Therefore, if $\bmu_{cl,s}$ can also be measured precisely, the hypothesis
of the cluster lens can be tested with high precision. 

The analyzed event was unusually sensitive to planets, 
independent of the possibility that the lens might be a cluster member. 
First, the source star is a low-luminosity giant, meaning that photometry
from both ground and space was unusually precise.   Second, it reached
magnification $A_\max\approx 13$ as seen from both Earth and \emph{Spitzer}.
Such moderate magnification events are substantially more sensitive
to planets than typical events \citep{mao91,gouldloeb}.  The combination
of these factors led to relatively intensive monitoring from the
ground and exceptionally intensive monitoring from \emph{Spitzer},
which further increased the event's planet sensitivity. 
We show that residuals for \emph{Spitzer} data and point-lens model 
can be fitted with a Saturn-mass ratio double-lens model. 
We do not claim planet detection because \emph{Spitzer} photometry 
of neighboring constant stars shows systematic trends that could mimic 
the planetary signal if superimposed on a purely point lens \citep{pac86} light curve. 
The only known planet in a globular cluster is in a system of white dwarf and pulsar \citep{richer03,nascimbeni12}. 

The light curve of OGLE-2015-BLG-0448 is analyzed here for two different purposes: 
to measure the microlens parallax and to estimate the planet sensitivity. 
The available ground-based data are survey observations by the Optical Gravitational Lens Experiment (OGLE) project 
and the follow-up observations taken by three groups: 
Microlensing Follow Up Network ($\mu$FUN), 
RoboNet, and 
Microlensing Network for the Detection of Small Terrestrial Exoplanets (MiNDSTEp).  
For parallax determination we use only OGLE photometry; Survey long-term monitoring 
is crucial in deriving event timescale and parallax. 
OGLE photometry is also well-characterized and systematic trends in the data are at a relatively low level. 
On the other hand, the planet sensitivity is the highest if many data points are taken close to the light curve maximum \citep{griest98}. 
The field including OGLE-2015-BLG-0448 is observed rarely by the OGLE survey, hence, 
the OGLE light curve does not contribute much to the planet sensitivity. 
The follow-up data give us much more information with this regard: 
they are taken from multiple sites allowing better time-coverage and reduced dependence on weather at a single site, and 
they can be also taken with much higher cadence because many telescopes are targeted on a single event. 
However, extending the event coverage by most of the follow-up observatories is not possible because of their limited resources 
or the chosen observing strategy. 
Additionally, many events get faint far from the peak and the smaller telescopes photometry in dense stellar region 
may be affected by systematic trends 
that could corrupt the measurements of the event timescale and parallax. 
Hence, the ground-based measurements of the event timescale and parallax are best done with the OGLE data only, 
but follow-up observations are included for the planet sensitivity calculations. 

We describe the observations in Section~\ref{sec:obs}.  
In Section~\ref{sec:anal} we analyze first the ground-based light curve alone and 
then the combined \emph{Spitzer} 
and ground-based light curves.  
We measure the microlens parallax $\bpi_\e$
and the closely related relative velocity projected on the observer plane $\tilde\bv_\hel$,
which are required to determine the lens location.  In Section~\ref{sec:pm}, 
we measure the proper motions of NGC 6558 and of the source star in the 
same frame of reference, which allows us to determine their relative
proper motion, $\bmu_{cl,s}$. The fact that its direction is inconsistent with that
of $\tilde\bv_\hel$ proves that the lens is not in the cluster.  Having eliminated
this possibility, in Section~\ref{sec:bulge}, we
demonstrate that the lens (host star) almost
certainly lies in the Galactic bulge, implying that it is a low-mass
star and that the tentative planet would  therefore be a cold Neptune.
The planet sensitivity of the event, which
will eventually be required for the determination of the Galactic
distribution of planets is analyzed in Section~\ref{sec:sens}.  We conclude
in Section~\ref{sec:conclude}. 
We discuss the tentative planet detection in Appendix~\ref{sec:planetApp}.

{\section{Observations}
\label{sec:obs}}

\subsection{OGLE Alert and Observations}

On 2015 March 20, the OGLE survey 
alerted the community to a new microlensing event OGLE-2015-BLG-0448
based on observations with the 1.4 deg$^2$ camera on  
the 1.3m Warsaw Telescope at the Las Campanas Observatory in Chile \citep{ogleiv}
using its Early Warning System (EWS) real-time event detection
software \citep{ews1,ews2}.  Most OGLE observations were taken in the $I$ band, 
and $V$ band observations are only used to determine the source properties. 
At equatorial coordinates ($18^{\rm h}10^{\rm m}14\fs38,~-31\degr45\arcmin09\farcs4$),
Galactic coordinates $(0\fdg20,\,-6\fdg01)$, this
event lies in the OGLE field BLG573, implying that it is observed
at roughly once per two nights (cf. Fig.~15 from \citealt{ogleiv}). 
We analyze 65 datapoints collected during the 2015 bulge season before 
${\rm HJD}^\prime \equiv {\rm HJD}-2450000  = 7301.6$ (Oct $6^{\rm th}$) 
and supplement them with 73 datapoints from 2014. To account for 
underestimated uncertainties that are reported by the image-subtraction
software we multiplied the uncertainties by a factor of $1.8$, 
so that the point-lens parallax model results in $\chi^2/dof \approx 1$.

\subsection{\emph{Spitzer} Observations}

OGLE-2015-BLG-0448 was announced by the \emph{Spitzer} team as a target
on 2015 May 19 UT 20:45 (${\rm HJD}^\prime = 7162.4$), 
about 2.5 weeks before the beginning of the
2015 \emph{Spitzer} observations (proposal ID: 11006, PI: Gould) 
and 3.5 weeks before this particular 
object could be observed (${\rm HJD}^\prime = 7187.1$) due to Sun-angle restrictions.
The reason for this early alert was that the source was bright and
appeared to be heading for relatively high magnification, making it relatively
sensitive to planets.  According to the protocols of \citet{yee15},
planet detections (and sensitivity) can only be claimed for observations
after the \emph{Spitzer} public selection date (or if the event was
later selected ``objectively'', which was not possible for this
event due to low OGLE cadence). Furthermore, without a public alert,
the event would not have attracted attention for the intensive follow-up required 
to raise sensitivity to planets.  The \emph{Spitzer} cadence was set
at once per day, and this cadence was followed during the second
week of the campaign, when OGLE-2015-BLG-0448 came within \emph{Spitzer}'s
view.

However, the \citet{yee15} protocols also prescribe that once all specified
observations are scheduled, any additional time should be allocated
to events that are achieving relatively high magnification during the
next week's observing window, with the cadence of these events
rank-ordered by the $1\,\sigma$ lower limit of expected magnification.
Based on this, OGLE-2015-BLG-0448 was slated for cadences of 4, 8, 8,
and 4 per day during weeks 3, 4, 5, and 6, respectively. Due to the fact
that it lay far to the east, OGLE-2015-BLG-0448 could be observed right
to the end of the campaign at ${\rm HJD}^\prime =7222.78$.  Altogether we collected 
210 epochs, each consisting of six 30s dithers. The photometry was obtained 
with a modified version of \citet{calchinovati15b} pipeline, which fits the centroid and brightness 
of every stars for each frame separately. The errorbars reported by this pipeline are a nearly 
linear function of the measured flux, hence, we assumed the errorbars are equal to the value of this linear 
function multiplied by the factor $4.3$ that brings $\chi^2/dof$ to 1 for the parallax point source model. 

\subsection{$\mu$FUN Observations}

As one of the few very bright \emph{Spitzer} events, and one that was
not intensively monitored by microlensing surveys 
(and so required follow-up to achieve reasonable planet sensitivity),
OGLE-2015-BLG-0448 was targeted by $\mu$FUN, 
including the following five small-aperture telescopes 
from Australia and New Zealand:
the Auckland Observatory 0.5m (R band), the Farm Cove Observatory 0.36m 
(unfiltered, Pakuranga), the PEST Observatory 0.3m (unfiltered, Perth),
the Possum Observatory 0.36m (unfiltered, Patutahi), and
the Turitea Observatory 0.36m (R band, Palmerston North). 
$\mu$FUN also observed the event regularly using the dual ANDICAM optical/IR
camera on the 1.3m SMARTS telescope at CTIO, Chile.  Almost all the optical
observations are in the $I$ band.  The IR observations are all in $H$ but 
these are for source characterization and are not used in the fits.
Follow-up photometric data were also taken by the Wise Collaboration on their 1.0m telescope at Mitzpe Ramon, Israel. 
A limited number of additional measurements were taken using two 0.7m 
MINiature Exoplanet Radial Velocity Array (MINERVA) telescopes at Mt. Hopkins, USA \citep{swift15}. 

All $\mu$FUN data were reduced using DoPhot software \citep{dophot}.  
The photometry of this event is hampered by 
an ab-type RR Lyrae variable OGLE-BLG-RRLYR-14873 \citep{kunder08,soszynski11} 
that lies projected at $2\farcs4$ from the event (Fig.~\ref{fig:find}), 
has $I$-band amplitude of $0.23\,{\rm mag}$, and period of $0.67\,{\rm d}$.  
Because 
DoPhot fits separately for the flux of each star at each epoch, it is
ideally suited to remove the effects of this neighboring variable,
even when the point spread functions (PSFs) of the two stars overlap, as they
frequently do for the smaller $\mu$FUN telescopes.  By contrast,
plain vanilla image-subtraction algorithms fit only for variations
centered at the source and so include residuals from neighboring PSFs,
if these overlap.  Unfortunately, DoPhot failed to separately identify
the source in PEST data and so these could not be used.  Possum data
showed unusual scatter and were also excluded.

\subsection{RoboNet Observations}

RoboNet 
observed OGLE-2015-BLG-0448 from three Las Cumbres Observatory Global Telescope Network (LCOGT) 
sites in its southern hemisphere ring of 1.0m telescopes: 
CTIO/Chile, SAAO/South Africa, and Siding Spring/Australia \citep{Brown2013}. 
Different telescopes at the same site are indicated as A, B, and C. 
Two CTIO telescopes (A and C) were equipped with the new generation of Sinistro imagers that consist of $4{\rm k}\times4{\rm k}$ Fairchild CCD-486 Bl CCDs 
and offer a field of view of $27^\prime\times 27^\prime$.  
Other telescopes support SBIG STX-16803 cameras with Kodak KAF-16803 front illuminated $4{\rm k}\times4{\rm k}$ pix CCDs, 
used in bin $2\times2$ mode with a field of view of $15.8^\prime\times 15.8^\prime$. 
All observations were made using SDSS-$i^\prime$ filters. 
Standard debiasing, dark-subtraction, and flat fielding were performed for all datasets by the LCOGT Imaging Pipeline, 
after which Difference Image Analysis was conducted using the RoboNet Pipeline, 
which is based around DanDIA \citep{Bramich2008, Bramich2013}.

LCOGT employed its TArget Prioritization algorithm \citep{Hundertmark2015} 
to select a sub-set of events from the \emph{Spitzer} target list based on their predicted sensitivity to planets, 
which were drawn from \emph{Spitzer} targets that  fell in regions of lower survey observing cadence.  
OGLE-2015-BLG-0448 was given priority because it fell within such a region, 
and due to the added scientific value of the proximity of the globular cluster.

\subsection{MiNDSTEp Observations}

The MiNDSTEp consortium observed OGLE-2015-BLG-0448 using the Danish 1.54 m telescope at ESO’s La Silla Observatory, Chile and 
the 0.35m Schmidt-Cassegrain telescope at Salerno University Observatory, Italy. 
The Danish telescope provides two-colour Lucky Imaging photometry using 
an instrument consisting of two Andor iXon+ 897 EMCCDs with a dichroic splitting 
of the signal at $655\,{\rm nm}$ into a red and a visual part, 
thereby collecting light from $466\,{\rm nm}$ to $655\,{\rm nm}$ (``extended $V$'') in the visual camera 
and from $655\,{\rm nm}$ to approximately $1050\,{\rm nm}$ (``extended $Z$'') in the red sensitive camera. 
The camera covers a $45''\times45''$ field of view on the $512 \times 512$ pixel EMCCDs with a scale of 0.09 arcsec/pixel and 
were operated at a frame rate of $10\,{\rm Hz}$ and a gain of 300~e$^-$/photon. 
On-line reductions and off-line re-reductions were performed with the Odin software \citep{Skottfelt15}, 
which is based on the DanDIA image subtraction and empirical PSF fitting. 
The Salerno data were taken in the $I$ band with a SBIG ST-2000XM CCD, 
and the images were reduced using a locally developed PSF fitting code. 
In total the Danish telescope has reported 148 $V$-band and 182 $Z$-band data points, 
and the Salerno University telescope 98 data points to the light curve of OGLE-2015-BLG-0448 
with the data collection strategy informed and implemented by means of the ARTEMiS system 
\citep[Automated Terrestrial Exoplanet Microlensing Search][]{dominik08}. 

We phased the residuals from the preliminary model with the pulsation period of the nearby RR Lyr and found 
significant contamination in the case of Salerno as well as LCOGT CTIO A and SSO B data. 
To correct for this contamination, we decomposed each of these datasets into 
source flux, blending flux, and scaled OGLE light curve of the RR Lyr. 
The contribution of the RR Lyr was then subtracted. 
Errorbars for every follow-up dataset were scaled so that $\chi^2/dof \approx 1$.

{\section{Lightcurve Analysis}
\label{sec:anal}}

We begin by fitting a simple five parameter model: $(t_0,u_0,t_\e,\bpi_\e)$
to the OGLE data. 
Here $(t_0,u_0,t_\e)$ are the standard \citet{pac86}
parameters, i.e., time of maximum light, impact parameter (scaled to
$\theta_\e$), and Einstein timescale, all as seen from Earth.  The
remaining two parameters are the microlens parallax vector $\bpi_\e$
\begin{equation}
\bpi_\e \equiv {\pi_\rel\over\theta_\e}\,{\bmu_\geo\over\mu_\geo};
\qquad
t_\e = {\theta_\e\over \mu_\geo},
\label{eqn:piedef}
\end{equation}
where $\theta_\e$ is the angular Einstein radius
\begin{equation}
\theta_\e^2 \equiv \kappa M\pi_\rel;
\qquad
\kappa \equiv {4 G\over c^2\,\au}\simeq 8.14\,{\mas\over M_\odot},
\label{eqn:thetaedef}
\end{equation}
$M$ is the lens mass, and $\pi_\rel\equiv\au(D_L^{-1}-D_S^{-1})$ and $\bmu_\geo$
are the lens-source relative parallax and proper motion, respectively,
the latter in the geocentric frame at the peak of the event as seen from the ground. 

Ground-based parallax models suffer from a two-fold degeneracy in $u_0$ \citep{smith03}. 
Table~\ref{tab:uparmO} presents parameters of the models with $u_0>0$ and $u_0<0$ 
that have almost the same $\chi^2$. We note that both models have similar $\pi_{{\rm E},E}$ 
but slightly different $\pi_{{\rm E},N}$, and $\pi_{{\rm E},N}>0$ 
at $2.2\sigma$ level. The fit to the OGLE data without parallax is worse by $\Delta\chi^2=10$.

\begin{figure}
\plotone{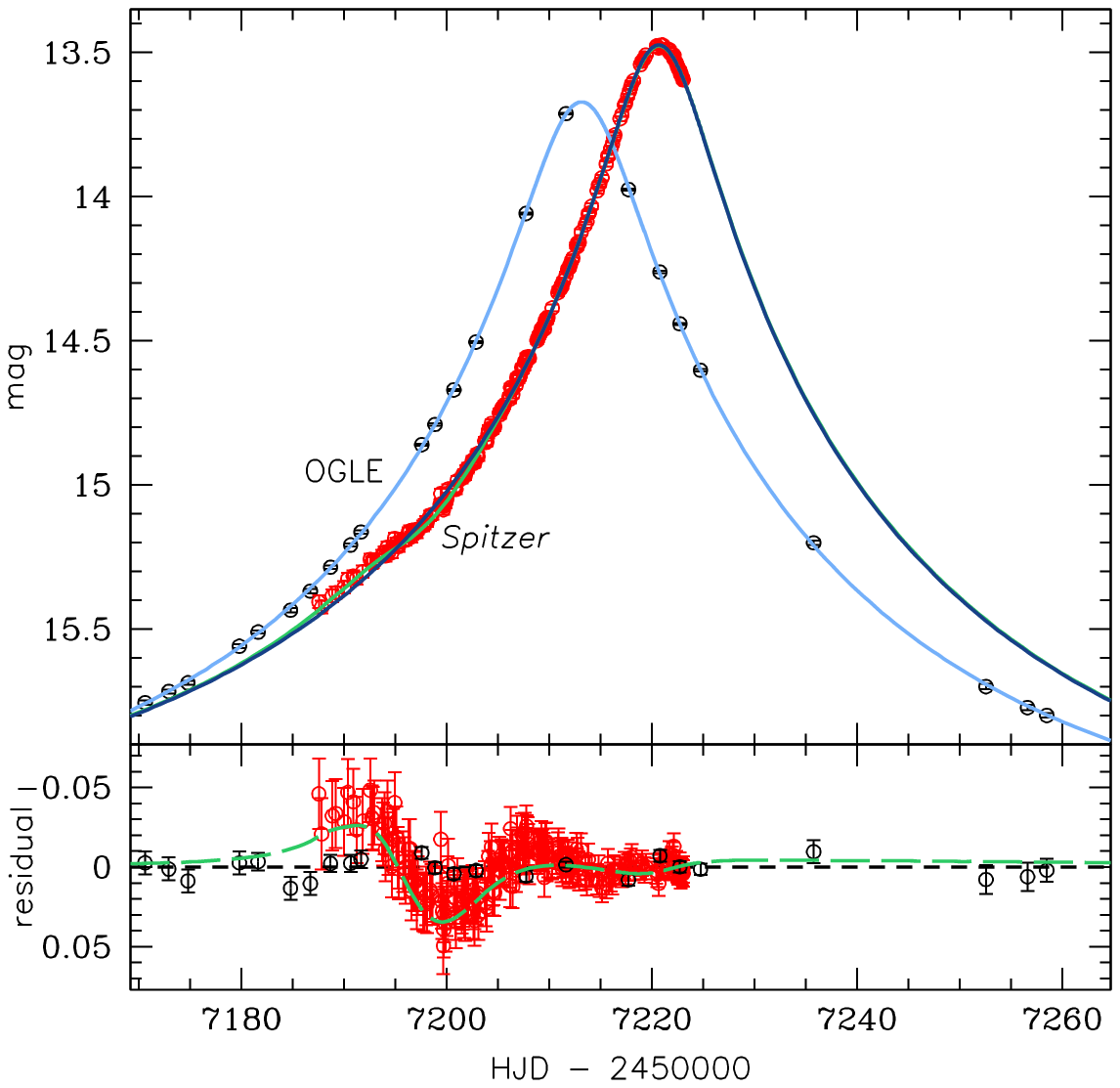}
\caption{Point-lens fit (with parallax) to \emph{Spitzer} and 
OGLE light curves of OGLE-2015-BLG-0448.  
The model (light blue line) fits the OGLE data (black points) quite well, 
but there are strong residuals in the \emph{Spitzer} data (red points and dark blue line),
particularly near the start of the observations.
The green line shows the planetary lens model for the \emph{Spitzer} data, which is discussed in Appendix~\ref{sec:planetApp}. 
The green long-dashed line in the lower plot shows the difference between the \emph{Spitzer} point-lens and double-lens models. 
}
\label{fig:lc}
\end{figure}

After fitting the OGLE data with a point-lens model, we analyze the OGLE and \emph{Spitzer} data jointly. 
The parallax point-lens fit (Figure~\ref{fig:lc}) 
shows significant systematic residuals in
the \emph{Spitzer} but not in the OGLE data.  Such a possibility was anticipated
by \citet{gouldhorne}, who suggested that space-based parallax observations
might uncover planets that are not detectable from the ground because
the spacecraft probes a different part of the Einstein ring.   However, it 
has never previously been observed.

\begin{figure}
\plotone{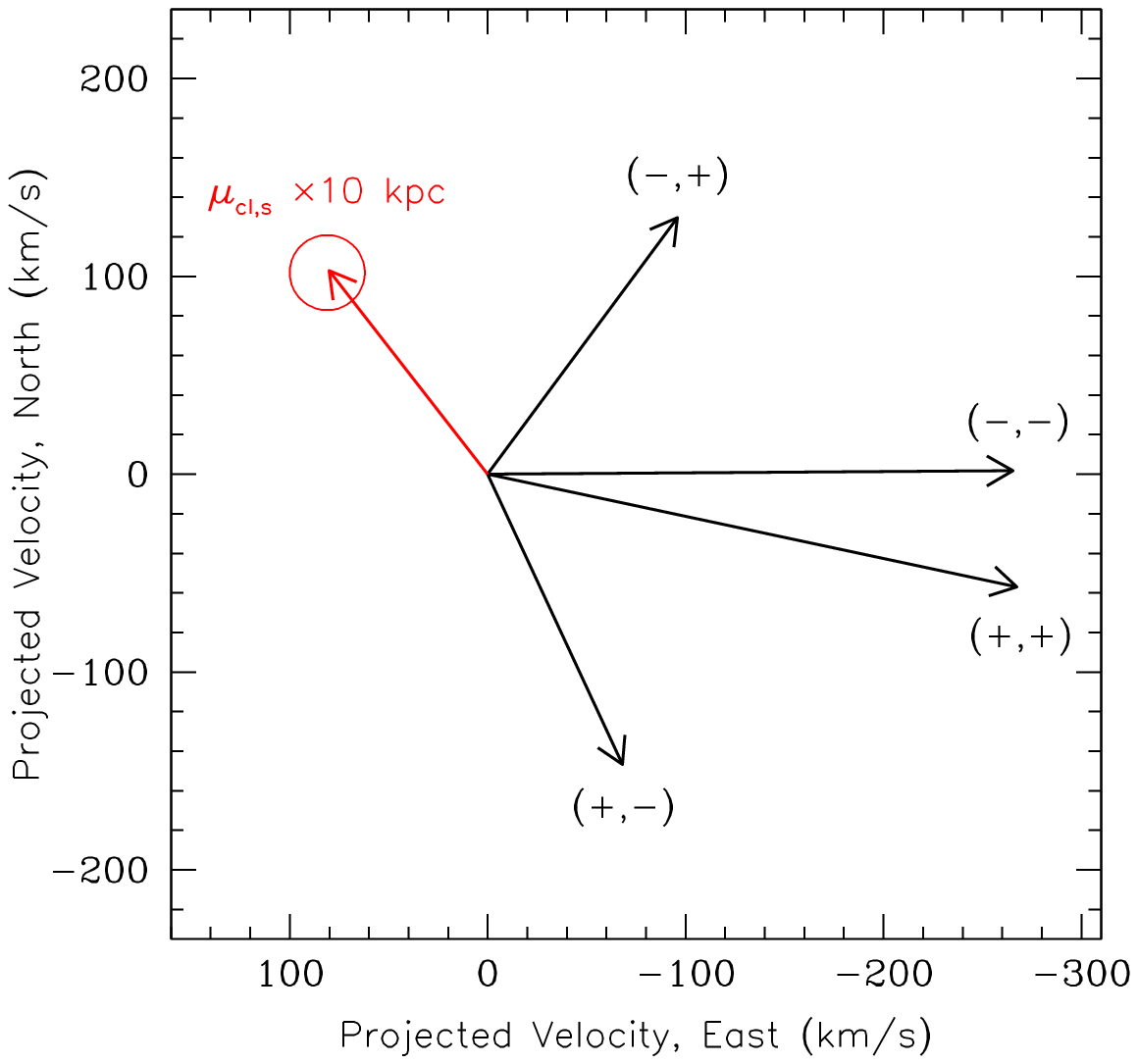}
\caption{Comparison of directions of astrometrically measured
$\bmu_{s,cl}$ (red) with four degenerate projected velocities $\tilde\bv_\hel$
based on microlensing data.  The proper motion measurement has been
scaled by an arbitrary distance (10 kpc) so that it has the same units
and approximately same size as the projected velocities. The 
direction of $\bmu_{s,cl}$ is inconsistent with any of the four $\tilde\bv_\hel$.
Hence, the lens does not belong to the cluster.
}
\label{fig:vt}
\end{figure}

The \emph{Spitzer} residuals are qualitatively similar to those analyzed
by \citet{gaudi02} for OGLE-1999-BUL-36. They found that this form of residuals 
could be explained either by a low mass-ratio companion ($q\ll 1$) with
projected separation (normalized to $\theta_\e$) $s<1$, or by light
curve distortions induced by the accelerated motion of the observer on
Earth, i.e., orbital parallax \citep{gould92}.  However, in the present
case, the latter explanation is ruled out because the parallax is
measured (and already incorporated into the fit) from the offsets in the
observed $(t_0,u_0)$ as seen from Earth and \emph{Spitzer},
$$ \bpi_{\e,\pm\pm} \simeq {\au\over D_\perp}(\Delta\tau,\Delta\beta_{\pm\pm}); $$
\begin{equation}
\qquad \Delta\tau \equiv {t_{0,\oplus} - t_{0,\rm sat}\over t_\e}; \\
\label{eqn:pieframe}
\end{equation}
$$ \qquad \Delta\beta_{\pm\pm} \equiv \pm u_{0,\oplus} - \pm u_{0,\rm sat}.$$
Here, ${\bf D}_\perp$ is the Earth-satellite separation projected 
on the sky (changes from $0.84$ to $1.31~{\rm AU}$ over the course of \emph{Spitzer} observations) and
where the subscripts $\oplus$ and ``sat'' indicate parameters as measured from Earth
and the satellite, respectively.  
The four solutions are specified $(\pm\pm)$ according to the signs of $u_0$ as seen
from Earth and \emph{Spitzer} respectively.  See \citet{gould04} for sign conventions. 
Table~\ref{tab:uparm} lists four possible solutions, including
the heliocentric projected velocity,
\begin{equation}
\tilde\bv_\hel = \tilde \bv_\geo + \bv_{\oplus,\perp};
\qquad
\tilde \bv_\geo = {\bpi_\e\over \pi_\e^2}\,{\au\over t_\e},
\label{eqn:vt}
\end{equation}
where $\bv_{\oplus,\perp}(N,E)= (-0.6,28.3)\,\kms$ is the velocity of Earth
projected on the sky at the peak of the event. 
The $(-+)$ solution is preferred over the other ones by $\Delta\chi^2=6.7$ 
because OGLE data prefer $\pi_{{\rm E},N} > 0$ 
and this solution has the highest $\pi_{{\rm E},N}$. 
The comparison of Tables~\ref{tab:uparmO} and \ref{tab:uparm} shows 
that the OGLE parallax measurement (that is based on slight light curve distortion) 
is consistent with the OGLE+\emph{Spitzer} result 
(that is based on the difference in $t_0$ and $u_0$ between the two observatories). 
Figure~\ref{fig:vt} displays the projected velocity vectors
for these four solutions.

\begin{table}
\caption{\label{tab:uparmO}
\sc OGLE-2015-BLG-0448 Point-Lens Parameters 
based on OGLE data} 
\vskip 1em
\begin{tabular}{@{\extracolsep{0pt}}llrr}
\hline
\hline
Parameter & Unit & $u_0>0$ & $u_0<0$ \\
$\chi^2$ &  & $ 125.1 $ &$ 125.0 $ \\
\hline
$t_0$ & day  & $ 7213.153 $ & $ 7213.153 $ \\
 &  & $\pm 0.016 $ & $\pm 0.016 $ \\
\hline
$u_0$ &   & $ 0.0874 $ & $ -0.0876 $ \\
 &  & $\pm 0.0016 $ & $\pm 0.0017 $ \\
\hline
$t_{\rm E}$ & day  & $ 61.23 $ & $ 60.83 $ \\
 & & $\pm0.84$ & $\pm0.95$ \\
\hline
$\pi_{{\rm E},N}$ &  & $ 0.113 $ & $ 0.180 $ \\
 & & $\pm 0.052 $ & $\pm 0.081 $ \\
\hline
$\pi_{{\rm E},E}$ & & $ -0.104$ & $ -0.111 $ \\
 & & $\pm 0.034 $ & $\pm 0.037 $ \\
\hline
$(F_b/F_{\rm base})_{\rm OGLE}$ &  & $ -0.002 $ & $ -0.004 $  \\
 & & $\pm 0.019 $ & $\pm 0.020 $ \\
\hline
\hline

\end{tabular}
\end{table}

There are only three possible causes of \emph{Spitzer} point-source
point-lens residuals: a binary (or planetary)
companion to the lens, a binary companion to the source, or an unmodeled systematics
in the light curve.  
Binary-source explanations for the residuals are basically ruled out
by the fact that no sign of source binarity is seen in the OGLE light curve.
Of course, one possible explanation for the lack of binarity effects would
be an extremely red source, which has so much less flux in $I$-band than
in \emph{Spitzer}'s $3.6\mu$m that it simply does not show up in the OGLE data.  
However, the 
source is a red giant, so there are very few stars on the color-magnitude
diagram (CMD) that are significantly redder.  For two of the solutions
($++$ and $--$) in Table~\ref{tab:uparm}, the source follows the
same trajectory as seen from Earth and \emph{Spitzer}, just separated
in time.  Hence, binary-source solutions are obviously inconsistent
with the OGLE data.  For the other two solutions, the second source
could pass farther from the lens as seen from Earth compared
to the \emph{Spitzer} by a factor 
$\approx 1+ (u_{0,\rm sat}+u_{0,\oplus})/u_{0,\rm sat^\prime}\approx 1+ 0.16/u_{0,\rm sat^\prime}$
where $u_{0,\rm sat^\prime}$ is the impact parameter of the source's companion
as seen by the \emph{Spitzer}. Given the slow development of the deviation,
$u_{0,\rm sat^\prime}\ga 0.1$, implying that this ratio of impact parameters is
$\la 2.6$. The source is already close to the reddest stars 
on the CMD, hence, 
the amplitudes of deviation have to be similar to the ratio of impact parameters,
which is clearly ruled out by the data.
Notwithstanding these general arguments, we fit for binary-source 
solutions.  We confirm that they are not viable. 
The binary lens models with planetary mass ratio are discussed in 
Appendix~\ref{sec:planetApp}. 
\newline

{\section{Proper Motion Measurements}
\label{sec:pm}}

\subsection{NGC 6558 Proper Motion Measurements in Literature}

The first measurement of the NGC 6558 proper motion was presented by 
\citet{vasquez13}. Stars on the upper red giant branch ($I < 16.5\,{\rm mag}$)
and bluer than bulge giants were selected as cluster members and the
mean proper motion of these stars was reported:
$\bmu_{cl}(N,E) = (0.06\pm0.14,0.52\pm0.14)\,\masyr$.
The bluer red giants were chosen because the metallicity of
the cluster stars is lower than the bulge red giants.  Hence, cluster
members on the giant branch are expected to be bluer. However,
the bulge red giants show significant metallicity spread
\citep{zoccali08}
and thus some bulge red giants can be mistaken for cluster members.
Therefore, one expects the \citet{vasquez13} measurement to be biased
toward smaller proper motion values. Additionally, the cluster proper
motion relative to the bulge could be underestimated because 
cluster members may have been included in the ensemble used to 
establish the ``bulge'' frame.

\citet{rossi15} published the only other NGC 6558 proper motion: 
$\bmu_{cl}(N,E) = (0.47\pm0.60,-0.12\pm0.55)\,\masyr$. 
In their approach cluster member selection and frame alignment
(needed for any proper motion measurement) were combined into one iterative
process. 
The CMD decomposed into cluster and field stars can be used to 
diagnose the reliability of this process. 
The most prominent cluster
feature on the CMD is the blue horizontal branch
defined by the stars of $V > 16$ and $(V-I) < 0.9$.
The decomposed CMDs for the cluster and the field reveal a very
similar number of stars in this region, even though we do not expect
field stars with these properties. The problems with decomposing blue
horizontal branch stars suggests that the iterative process used to
select cluster members and measure proper motions, failed in this case.

\subsection{NGC 6558 Proper Motion Measurement From OGLE-IV Data}

We use two different methods to measure the proper motion of NGC 6558.
In both cases, we make use of 5 years of OGLE-IV observations of this
field.  We first establish a ``Galactic bulge reference frame'' by 
identifying red giant stars from the CMD on the
chip where the cluster lies, but excluding a circle of radius $1\farcm52$
around the cluster itself\footnote{The
NGC 6558 cluster core radius and half-light radius are $0\farcm03$ and 
$2\farcm15$, respectively. 
The cluster tidal radius is $10^{2.50}$ times the core radius \citep[][2010 edition]{harris96}.
OGLE-2015-BLG-0448 lies $58\farcs5$ from the center.}.
 We note that for the immediate purpose
of this paper, it is not important whether this reference frame is
contaminated by non-bulge stars because we will measure the proper motion
of the source in the same frame.  However, the general utility of
this measurement does require that this be the bulge frame, and the
red giants are the best way to define this. 
Because the reference frame is defined by 2000 stars whose dispersion
is about $2.7\,\masyr$ in each direction, it 
is randomly offset from the ``true bulge frame'' by $0.06\,\masyr$
in each direction.

In the first method, we measure the proper motion of each star 
$I<18\,{\rm mag}$ within a radius of $0\farcm87$ from the cluster center.
We fit the resulting distribution of 518 proper motion measurements
to the sum of two two-dimensional Gaussians, described by a total of
four parameters, i.e., the cluster proper motion $\bmu_{cl}$, a single
isotropic ``cluster'' dispersion $\sigma_{cl}$ (actually mostly
due to measurement error rather than intrinsic dispersion),
and the fraction of all
stars in the sample that belong to the cluster, $p$.  The second
Gaussian is assumed to have the same properties as the bulge population,
i.e, a centroid at $(0,0)$ and a dispersion $(2.7,2.7)\,\masyr$.

\begin{figure}
\includegraphics[bb=24 30 404 745,height=.7\textheight]{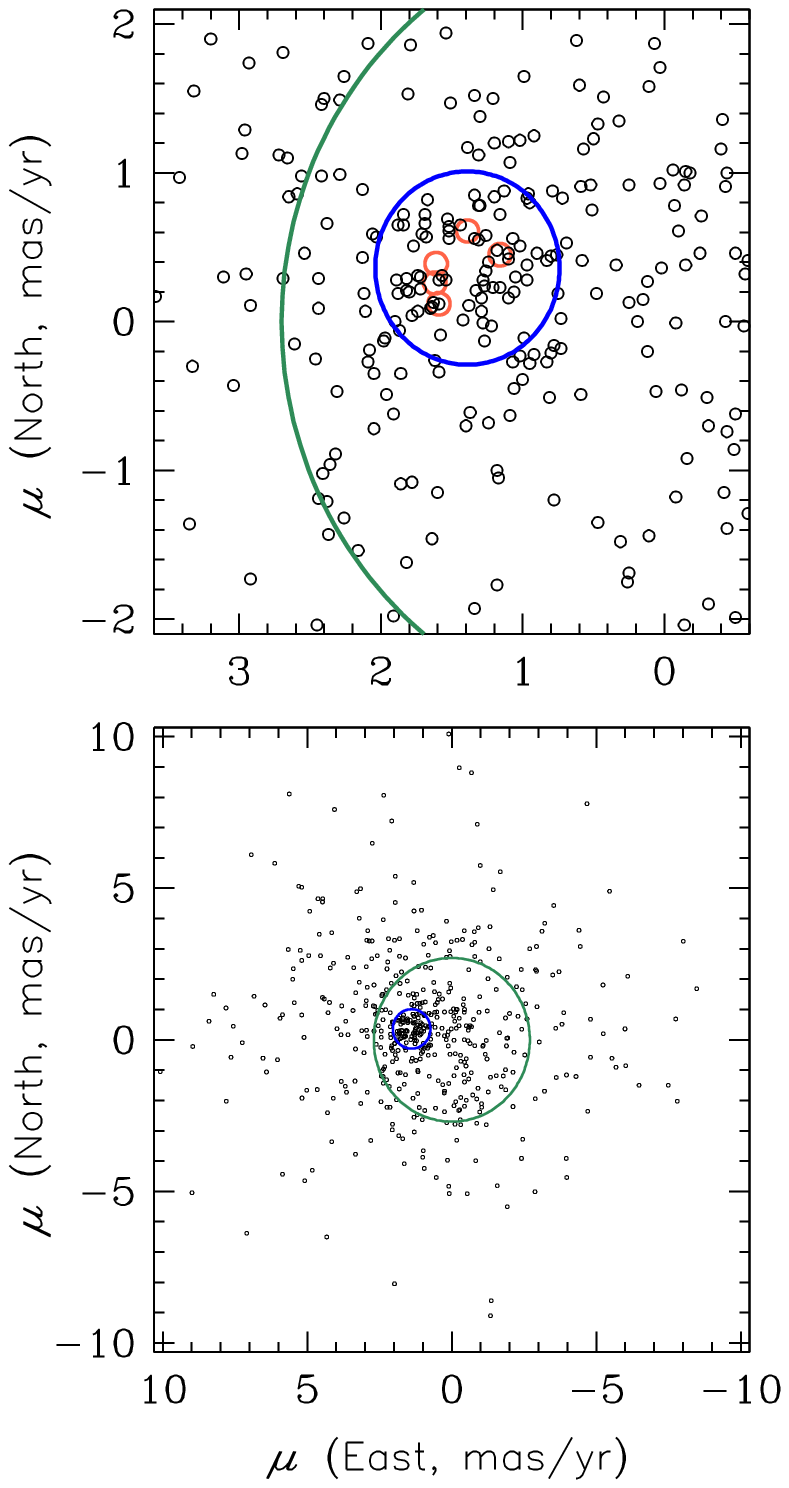}
\caption{Proper motions of stars within $0\farcm87$ of the center of
NGC 6558 based on OGLE-IV data.  The distribution was fit to the sum
of two Gaussians, one for the bulge, centered at (0,0) and with
the known bulge dispersion $\sigma=2.7\,\masyr$ (green circle), and
the other with freely fit center and dispersion (blue circle).
This gives one measure of the cluster proper motion in the bulge
frame $\bmu_{cl}(N,E) = (+0.36\pm 0.08,+1.39\pm 0.08)$. In a second
method, we take the average proper motion of five spectroscopically
confirmed cluster members (small red circles, upper zoomed panel only),
which yields $\bmu_{cl}(N,E) = (+0.33\pm 0.08,+1.49\pm 0.08)$.
Since these are consistent, we combine them to yield 
Equation~(\ref{eqn:clusterpm3}).
}
\label{fig:cl}
\end{figure}

We find $p=24\pm 3\%$, $\sigma_{cl}=0.65\,\masyr$, and
\begin{equation}
\bmu_{cl,1}(N,E) = (+0.36\pm 0.08,+1.39\pm 0.08)\,\masyr.
\label{eqn:clusterpm}
\end{equation}
See Figure~\ref{fig:cl}.

In the second method, we measure the proper motions of five spectroscopically
confirmed cluster members \citep{zoccali08,dias15}, and find
\begin{equation}
\bmu_{cl,2}(N,E) = (+0.37\pm 0.08,+1.47\pm 0.09)\,\masyr,
\label{eqn:clusterpm2}
\end{equation}
where the error is determined from the scatter. 
See the upper panel of Figure~\ref{fig:cl}.
Since these are consistent at the $1\,\sigma$, we combine the
two measurements to obtain
\begin{equation}
\bmu_{cl}(N,E) = (+0.36\pm 0.06,+1.42\pm 0.06)\,\masyr.
\label{eqn:clusterpm3}
\end{equation}
We remind the reader that these errors are relative to the frame,
which is what is relevant to our current application.  Since
the frame itself has errors of $0.06\,\masyr$, the total error
in this value in the ``true bulge frame'' is $0.08\,\masyr$.

\subsection{Proper Motion of Source Star}

We measure the proper motion of the OGLE-2015-BLG-0448 source in the same frame:
\begin{equation}
\bmu_{s}(N,E) = (-1.81\pm 0.40,-0.27\pm 0.40)\,\masyr.
\label{eqn:sourcepm}
\end{equation}
We estimate the error in two ways.  First, we note that the two methods
of measuring $\bmu_{cl}$ revealed scatters of $0.65\,\masyr$ and
$0.18\,\masyr$ for the two star samples with median brightness of $I\approx 17.2\,{\rm mag}$
and $I\approx 14.2\,{\rm mag}$, respectively.  Given that the OGLE-2015-BLG-0448
source has a baseline magnitude of $I_{\rm base}=16.34$, we adopt an intermediate
value of $0.40\,\masyr$.  Second, substantial experience from regions
where two OGLE fields overlap, shows that proper-motion errors are typically at
about this level for $I\approx 16.5\,{\rm mag}$ stars.

The relative proper motion between the cluster and the source-star is 
\begin{equation}
\bmu_{cl,s}(N,E) = (+2.17\pm 0.40,+1.69\pm 0.40)\,\masyr.
\label{eqn:clspm}
\end{equation}

\subsection{Lens Is Not Cluster Member}

We put the proper motion vector $\bmu_{cl,s}$ (Equation~(\ref{eqn:clspm}))
on Figure~\ref{fig:vt} in order to test whether its direction
is consistent with any of the lens-source projected velocities.
Because $\bmu_{cl,s}$ and $\tilde \bv_\hel$ have different units, $\bmu_{cl,s}$
must be multiplied by a dimensional quantity in order to be displayed
on the same plot.  We call this $D_\rel$ for reasons that will become clear.
We have chosen $D_\rel = 10\,\kpc$ simply because the vectors
are then roughly the same size.  
The $\bmu_{cl,s}$ is clearly inconsistent with any of the
four values of $\tilde \bv_\hel$, hence the lens is definitely not in the cluster.

However, if $\bmu_{cl,s}$ had been consistent with one of the
$\tilde \bv_\hel$, then the $D_\rel$ required to make the two
vectors in Figure~\ref{fig:vt} align would have provided an additional test
for cluster membership.  That is,
\begin{equation}
{\tilde v_\hel\over\mu_{l,s}} = {\au\over\pi_\rel} \rightarrow D_\rel,
\label{eqn:drel}
\end{equation}
where $\bmu_{l,s}$ is the lens-source relative proper motion, which
for our purposes can be taken as identical to the cluster proper motion, because 
$|\mu_{l,s}-\mu_{cl,s}|\leq |\bmu_{l}-\bmu_{cl}| = v_{l,cl}/D_L \la 0.2\,\masyr$.
Here $\bv_{l,cl}$ is the lens velocity in the cluster frame.

If, for example, $\bmu_{cl,s}$ had been in exactly the opposite direction to the one measured, 
it would have been consistent in direction with $\tilde\bv_{\hel,+-}$.
Then, identifying the lens as in the cluster would have implied
$\pi_{\rel,cl,s}\simeq 100\,\muas$.  This would have been an
implausible value because the cluster is believed to be at $D\approx 7\,\kpc$,
i.e, $\pi_{cl}\approx 140\,\muas$, which would imply $\pi_s = 40\,\muas$,
i.e., $D_L=25\,\kpc$.  That is, the $D_\rel$ required to align
$\bmu_{cl,s}$ and $\tilde \bv_\hel$ provides a powerful consistency
check on the identification of the lens as a cluster member.

{\section{Location of Lensing System}
\label{sec:bulge}}

For a large fraction of past planetary microlensing events, $\theta_\e$ is
measured from the finite source effects, since the model then yields
$\rho=\theta_*/\theta_\e$ and the angular source radius $\theta_*$
is easily measured \citep{ob03262}.  Unfortunately, this event contains
no caustic crossings or cusp approaches so this standard method cannot
be applied.  \citet{21event} showed that for events with measured
parallaxes $\bpi_\e$, the lens distance (and hence the mass) could be
estimated kinematically, with relatively small error bars.  However,
of the 21 events analyzed there, all but one had projected velocities
that either were quite large $\tilde v_\hel>700\,\kms$ or were consistent 
in direction with Galactic rotation.  The first group are easily
explained as Galactic bulge lenses $\pi_\rel\la 0.02\,\mas$, since
$\mu = \tilde v_\hel \pi_\rel/\au = 3\,\masyr 
(\tilde v/700\,\kms)(\pi_\rel/0.02\,\mas)$, which is a 
typical value for bulge lenses.  The second group are easily
explained as lenses rotating with the Galactic disk, with the magnitude
of $\tilde v_\hel$ giving a rough kinematic distance estimate
$\pi_\rel \simeq \mu_{sgrA*}\au/\tilde v_\hel$ and $\mu_{sgrA*}=6.38\,\masyr$
is the proper motion of SgrA*.  The one exception was OGLE-2014-BLG-0807,
for which the favored solutions had $\tilde v_\hel \approx 200\,\kms$. 

The best model $(-+)$ in Table~\ref{tab:uparm} has $\tilde v_\hel = 161.2(4)\,\kms$, 
while the models $(++)$ and $(--)$ that fit the data slightly worse, predict $\tilde v_\hel = 270\,\kms$. 
Neither of the $\tilde\bv_\hel$ vectors is aligned with Galactic disk rotation, 
hence there is a low probability that the lens is in the Galactic disk. 
The measured projected velocity could be explained by a bulge lens 
if the lens-source relative parallax were larger than typical. 
The line of sight toward the event at Galactic coordinates 
$(l, b) = (0\fdg20,\,-6\fdg01)$ crosses the two arms of 
the bulge X-shaped structure \citep{nataf10,mcwilliam10,gonzalez15}.
Hence, it is possible that the lens is in the closer part of the bulge 
and the source is much further away and the relative parallax is higher than typical. 
Even in this case the $\tilde v_\hel = 270\,\kms$ solutions would be preferred over 
$\tilde v_\hel = 160\,\kms$, i.e., contrary to the least-squares fits to the OGLE data. 
In either case, the most probable lens location is in the closer part of the bulge. 
\newline

{\section{Planet Sensitivity}
\label{sec:sens}}

With peak magnifications of 11 (from ground) and 14 (from \emph{Spitzer}), 
and average cadences of 36 per day (ground-based survey plus follow-ups) 
and 6 per day (for \emph{Spitzer}), 
event OGLE-2015-BLG-0448 is among the \emph{Spitzer} 2015 events 
that are most sensitive to planet perturbations. 
Therefore, we present the planet sensitivity of this event here, 
it will also be required for the determination of the Galactic distribution of planets, 
no matter whether the planet detection in this event is real or not.

We compute the planet sensitivity of this event using the method that 
was first proposed by \citet{rhie00} and further developed by 
\citet{yee15} and \citet{zhu15} to include space-based observations. 
Details of the method can be found in the latter two references. 
In brief, we first measure the planet sensitivity $S$ as a function of $q$ and $s$. 
For each set of $(q,s)$, we generate 300 planetary light curves that vary in 
angle between the source trajectory and  the lens binary axis, $\alpha$, 
but have other parameters fixed to the observed values. 
For each simulated light curve, we then find 
the best-fit single-lens model using the downhill simplex algorithm. 
The deviation between the simulated data and its best-fit single-lens 
model is quantified by $\chi^2_{\rm SL}$. For a subjectively chosen event, 
which is the case of OGLE-2015-BLG-0448, 
we first fit the simulated data that were released before the selection date 
$t_{\rm select}$ and find $\chi^2_{\rm SL,select}$. 
If $\chi^2_{\rm SL,select}>10$, we regard the injected planet as 
having been noticeable and thus reject this $\alpha$; otherwise 
we compare $\chi^2_{\rm SL}$ from the whole light curve with our 
pre-determined detection threshold, and consider the injected planet 
as detectable if $\chi^2_{\rm SL}>\chi^2_{\rm threshold}$. 
The sensitivity $S(q,s)$ is the fraction of $\alpha$ 
values for which the planet is detectable. 
We assume \"Opik's law in $s$, i.e., a flat distribution of $\log{s}$, 
and compute the integrated planet sensitivity $S(q)$.

We adopt the following detection thresholds, which are more realistic than that used in \citet{zhu15}:
C1: $\chi^2_{\rm SL}>300$ and at least three consecutive data points showing $>3~\sigma$ deviations; or
C2: $\chi^2_{\rm SL}>500$.
C1 is used mainly to recognize sharp planetary anomalies. Some of these anomalies might not be treated as reliable detections with only the current data, because of the low $\chi^2_{\rm SL}$. However, they are nevertheless significant enough to trigger the automatic anomaly detection software and/or attract human attentions, either of which would lead to dedicated follow-up observations of the anomalies and thus confirm these otherwise marginal detections. C2 as a supplement of C1 intends to capture the long-term weak distortions that may not show sharp deviations.

The calculation of planet sensitivity requires $\rho$ as an input. 
Here we estimate $\rho$ following the prescription given by \citet{yee15}: 
$\rho = \theta_*/\theta_\e$ where $\theta_\e = \pi_\rel / \pi_\e$. 
The parallax $\pi_\e$ is well measured thanks to a combination of the OGLE and the \emph{Spitzer} data, 
hence below we need to estimate only $\pi_\rel$ and $\theta_*$. 
The lens-source relative parallax can be easily found under the assumption 
that the lens is in the closer arm of the X-shaped structure and the source is in the further arm. 
We follow \citet{nataf15} who in detail modeled properties of the X-shaped structure in OGLE-III fields. 
The two centroids of RC luminosity functions corrected for extinction are 
$I_{{\rm RC}1,0} = 14.210~{\rm mag}$ and $I_{{\rm RC}2,0} = 14.715~{\rm mag}$ for the event location 
(average values for fields BLG169 and BLG170). 
For absolute RC brightness of $M_{I,{\rm RC}} = -0.12~{\rm mag}$ 
the corresponding distances are $7.3$ and $9.3~{\rm kpc}$, hence, $\pi_\rel = 0.028~{\rm mas}$. 

To calculate $\theta_*$ we assume the source $I$-band brightness and $(V-I)$ color 
are the same as the baseline object: $I_s = 16.337~{\rm mag}$ and $(V-I)_s = 1.589~{\rm mag}$ \citep{szymanski11}. 
This is justified because none of our models predicts significant blending. 
We corrected for extinction using \citet{nataf13} extinction maps and obtain: 
$I_{s,0} = 15.711~{\rm mag}$ and $(V-I)_{s,0} = 1.046~{\rm mag}$. 
This $(V-I)_{s,0}$ corresponds to $(V-K)_{s,0} = 2.419~{\rm mag}$ \citep{bb88}. 
The \citet{kervella04} color-surface brightness relation gives $\theta_* = 3.4~{\rm \mu as}$. 
Finally, $\rho = \theta_*\pi_\e/\pi_\rel = 0.019$ and $0.011$ for $(-+)$ and $(--)$ models, respectively. 

We plot all the ground-based data in Figure~\ref{fig:lc_followup}. 
The highest contribution to the planet sensitivity comes from the Auckland and LCOGT CTIO A datasets. 
We compute the planet sensitivity for two out of four possible solutions, 
$(--)$ and $(-+)$, and show the results in Figure~\ref{fig:sensitivity}.  
Both solutions show substantial planet sensitivity 
($>10\%$) down to $q=10^{-4}$. 
The $(-+)$ solution shows slightly higher sensitivity for $q \gtrsim 2\times10^{-4}$, 
mostly because observations taken from the satellite and Earth are probing different regions in the Einstein ring, 
as has been discussed in \citet{zhu15} 
and the reader can also see Figure~\ref{fig:chisq-map} here for a demonstration. 
At smallest $q$ values the $(-+)$ solution is less sensitive than the $(--)$ solution, 
because the larger source size ($\rho=0.019$) smears out subtle features due to small planets. 
Figure~\ref{fig:chisq-map} shows the detectability of planets with mass ratio 
$q=1.70\times10^{-4}$ as functions of planet positions for both investigated solutions. 
It is clear that the tentative planet detection reported here 
can only happen in the $(-+)$ solution.
\newline

{\section{Conclusions}
\label{sec:conclude}}

The event OGLE-2015-BLG-0448 presented a number of unique properties. 
It lay projected within tidal radius of the globular cluster. 
The maximum magnification reached was relatively high both for \emph{Spitzer} and ground-based observations. 
It was also intensively monitored both from the ground and from space. 
All these properties made it a potential probe of the population of planets 
in globular clusters. 

We analyzed the event photometry from both \emph{Spitzer} and ground-based telescopes: the OGLE survey and follow-up networks of $\mu$FUN, RoboNet, and MiNDSTEp. 
Microlens parallax was measured using the difference in event properties as seen from ground and space. 
The result confirmed the microlens parallax measured using only the OGLE data. 
Additionally, long-term astrometry of OGLE images was used to measure proper motions. 
We measured the proper motion of globular cluster NGC 6558 and the event source. 
Our analysis reveals that the lens could not be a cluster member. 
The same methods can be used for other potential cluster lens events that are observed by satellites. 

We found that the \emph{Spitzer} light curve reveals significant trends in residuals of the point-source point-lens model. 
The only two plausible causes of these trends are problems with \emph{Spitzer} photometry or a planetary companion to the lens.  
We do not claim planet detection, but provide results of planetary model fitting in case the event photometry is proven correct.

\acknowledgments

The OGLE project has received funding from the National Science Centre,
Poland, grant MAESTRO 2014/14/A/ST9/00121 to AU. 
Work by WZ and AG was supported by NSF AST 1516842.
Work by YS \& CBH was supported by an appointment to the NASA Postdoctoral Program at the Jet Propulsion Laboratory, 
administered by Oak Ridge Associated Universities through a contract with NASA. 
Work by JCY, AG, and SC was supported by JPL grant 1500811.  Work by JCY was
performed under contract with the California Institute of Technology
(Caltech)/Jet Propulsion Laboratory (JPL) funded by NASA through the
Sagan Fellowship Program executed by the NASA Exoplanet Science
Institute. 
This publication was made possible by NPRP grant \# X-019-1-006 from the Qatar National Research Fund (a member of Qatar Foundation). 
Work by SM has been supported by the Strategic Priority Research Program ``The Emergence of Cosmological Structures" of the Chinese Academy of Sciences Grant No. XDB09000000, and by the National Natural Science Foundation of China (NSFC) under grant numbers 11333003 and 11390372. 
MPGH acknowledges support from the Villum Foundation. 
This work makes use of observations from the LCOGT network, 
which includes three SUPAscopes owned by the University of St Andrews. 
The RoboNet programme is an LCOGT Key Project using time allocations from the University of St Andrews, 
LCOGT and the University of Heidelberg together with 
time on the Liverpool Telescope through the Science and Technology Facilities Council (STFC), UK. 
This research has made use of the LCOGT Archive, 
which is operated by the California Institute of Technology, 
under contract with the Las Cumbres Observatory. 
Operation of the Danish 1.54m telescope at ESO’s La Silla observatory was supported by The Danish Council for Independent Research, Natural Sciences, and by Centre for Star and Planet Formation.

\appendix
\section{Tentative Planet}
\label{sec:planetApp}

The point source model fitted to the \emph{Spitzer} data resulted in residuals 
with significant trends. Here we discuss the possibility that these 
residuals were caused by the companion to the lens. 

The only possible binary-lens solutions must have planetary mass ratios
$q\ll 1$ and separations satisfying 
$|s-s^{-1}|\approx 0.5$, i.e., $\log s\approx \pm 0.11$, which follows from
simple arguments.  
First, the source passes the lens at $u_0\approx 0.08$ as seen
from both Earth and \emph{Spitzer}.  Since neither light curve is
perturbed at peak, this already implies that the central caustic is
small.  Such small central caustics require either $s\ll 1$, $s\gg 1$,
and/or $q\ll 1$.  However, if either of the first two held, there
could not be a significant perturbation at the point that it is
observed at $u_{\rm sat}\approx 0.5$.  That is, the event timescale
$t_\e\approx 60\,$days is set by the unperturbed OGLE light curve.
Hence, the fact that the \emph{Spitzer} curve experiences an
excess roughly 30 days before peak implies that there is a caustic
structure at $u_{\rm sat}\approx 30/60=0.5$.  

Thus, $q\ll 1$.  In this planetary regime,
such caustics occur when the planet is aligned to one of the two
unperturbed images of the primary lens at $u=|s-s^{-1}|$, 
i.e., $s=|-u\pm(u^2+4)^{1/2}|/2$.  Hence, $|\log s|\approx 0.11$.

Finally, the fact that the \emph{Spitzer} light curve
is perturbed while the OGLE light curve is not, implies (as in
the above binary source analysis), that the source passes on opposite 
sides of the lens ($+-$ or $-+$ solutions). The preference of $(-+)$ in 
Table~\ref{tab:uparm} makes it the best solution. 

We consider four different topologies obeying the above constraints.
First, $s<1$ with the source (seen by \emph{Spitzer}) passing between
the two triangular caustics for this topology.  Second, $s<1$ with
the source passing outside one of these caustics.  Third, $s>1$.
For each topology, we insert a series of seed solutions as a function
of $q$ and allow all parameters to vary.  We find that the first
and the third topologies never match the observed morphology of the
\emph{Spitzer} light curve because their relative demagnification
zones do not align to the relative ``dip'' in the \emph{Spitzer}
light curve at about ${\rm HJD}^\prime =7200$.  The second topology always converges
to the same solution, which we present in Figure~\ref{fig:traj}. 
The model \emph{Spitzer} light curve is shown in Figure~\ref{fig:lc} by green line. 
The single lens parameters are consistent with the $(-+)$ solution in Table~\ref{tab:uparm}: $t_0=7213.161(14)$, $u_0=-0.0870(10)$, $t_\e=61.16(16)\,{\rm d}$, $\pi_{\e, N}=0.1140(12)$, $\pi_{\e, E}=-0.1088(10)$, and $F_b/F_{\rm base, OGLE} = 0.002(11)$. 
The additional binary lens parameters are: $\alpha=189\fdg71(25)$, $s=0.7870(50)$, and $q=1.70(32)\times10^{-4}$. 
The $\chi^2/dof = 209.7/331$ 
is better by $\chi^2 = 127.7$ than the point-lens solution, and better by $\Delta\chi^2 = 49$ 
than the double-lens $(+-)$ solution. 
We note that even the best-fitting model does not remove all the systematics seen in \emph{Spitzer} residuals. 

The light curve lacks close approach to the caustics, 
which is uncommon among published microlensing planets \citep{zhu13}. 
Without the caustic approach we are unable to constrain the source size relative to $\theta_{\rm E}$. 
We note that \citet{yee13} found a planetary signal below the reliability 
threshold in MOA-2010-BLG-311 event that also lies close to a globular cluster (NGC 6553 in that case).

\begin{table}[p]
\caption{\label{tab:uparm}
\sc OGLE-2015-BLG-0448 Point-Lens Parameters based on OGLE and \emph{Spitzer} data}
\vskip 1em
\begin{tabular}{@{\extracolsep{0pt}}llrrrr}
\hline
\hline
Parameter & Unit & $(++)$ & $(--)$ & $(+-)$ & $(-+)$ \\
\hline \hline
$\chi^2$ & &$ 346.5 $ &$ 344.1 $ &$ 380.3 $ &$ 337.4 $ \\
\hline
$t_0$ & day &$ 7213.135 $ &$ 7213.136 $ &$ 7213.116 $ &$ 7213.146 $ \\
 & &$\pm 0.014 $ &$\pm 0.014 $ &$\pm 0.014 $ &$\pm 0.014 $ \\
\hline
$u_0$ &  &$ 0.0863 $ &$ -0.0866 $ &$ 0.0853 $ &$ -0.0874 $ \\
 & &$\pm 0.0010 $ &$\pm 0.0010 $ &$\pm 0.0010 $ &$\pm 0.0010 $ \\
\hline
$t_{\rm E}$ & day  &$ 61.91 $ &$ 61.68 $ &$ 62.51 $ &$ 61.02 $ \\
 & &$\pm 0.51 $ &$\pm 0.51 $ &$\pm 0.52 $ &$\pm 0.51 $ \\
\hline
$\pi_{\rm E,N}$ &  &$ -0.0174 $ &$ 0.0008 $ &$ -0.1321 $ &$ 0.1142 $ \\
 & &$\pm 0.0005 $ &$\pm 0.0005 $ &$\pm 0.0014 $ &$\pm 0.0012 $ \\
\hline
$\pi_{\rm E,E}$ &  &$ -0.0912 $ &$ -0.0956 $ &$ -0.0870 $ &$ -0.1088 $ \\
 & &$\pm 0.0009 $ &$\pm 0.0009 $ &$\pm 0.0008 $ &$\pm 0.0010 $ \\
\hline
$(F_b/F_{\rm base})_{\rm OGLE}$ &  &$ 0.013 $ &$ 0.009 $ &$ 0.026 $ &$ -0.002 $ \\
 & &$\pm 0.011 $ &$\pm 0.011 $ &$\pm 0.011 $ &$\pm 0.011 $ \\
\hline
\hline
$\tilde v_{\rm N,hel}$ & $\rm km\,s^{-1}$  &$ -56.93 $ &$ 1.77 $ &$ -146.84 $ &$ 129.67 $ \\
 & &$\pm 1.30 $ &$\pm 1.22 $ &$\pm 0.41 $ &$\pm 0.36 $ \\
\hline
$\tilde v_{\rm E,hel}$ & $\rm km\,s^{-1}$ &$ -267.54 $ &$ -265.28 $ &$ -67.96 $ &$ -95.78 $ \\
 & &$\pm 0.80 $ &$\pm 0.85 $ &$\pm 0.37 $ &$\pm 0.41 $ \\
\hline
\hline

\end{tabular}
\end{table}

\begin{figure}[p]
\begin{center}
 \includegraphics[width=.75\textwidth]{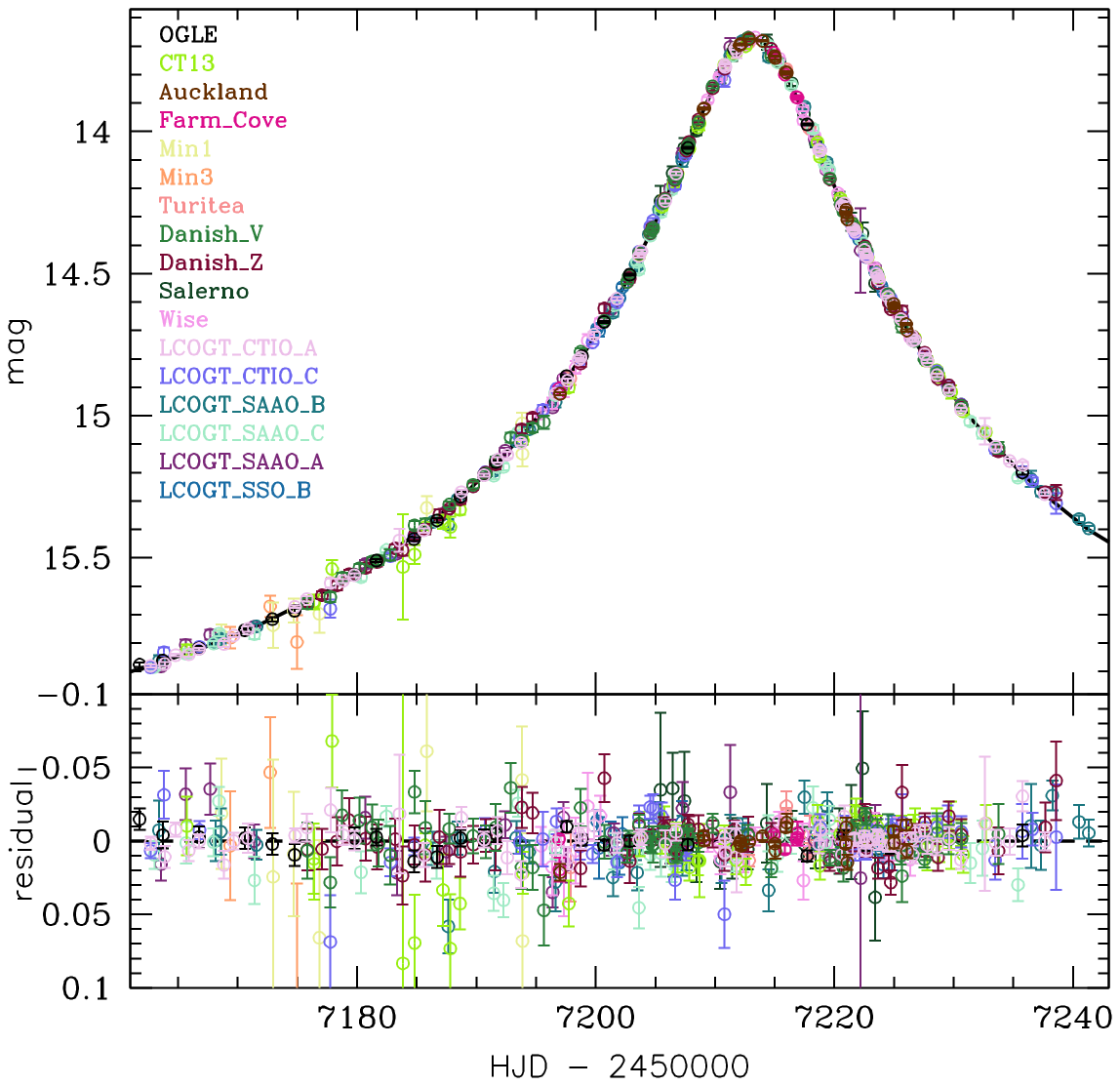}
\end{center}
\caption{
Ground-based light curve of OGLE-2015-BLG-0448. Different colors represent different datasets. 
For clarity, the follow-up data were averaged in bins separately chosen for each dataset. 
The bins were set based on comparison of the uncertainty of the mean point and the change of the model brightness over the bin timespan. 
For each bin the uncertainty of the mean point is smaller than the maximum difference between the model brightness and the mean model value. 
There are 462 bins that are based on 1638 follow-up data points. 
\label{fig:lc_followup}}
\end{figure}

\begin{figure}
\includegraphics[width=.81\textwidth]{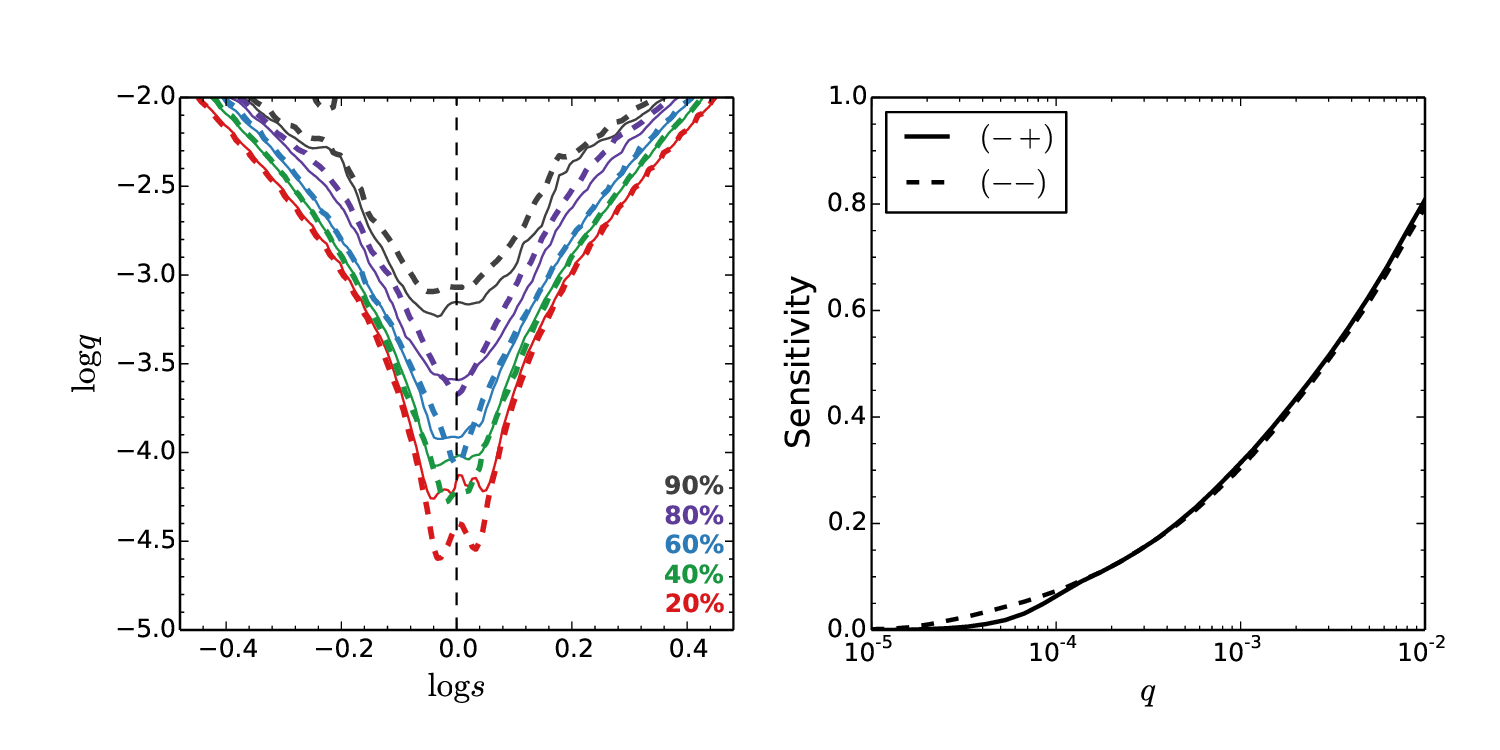}
\caption{Planet sensitivity results of OGLE-2015-BLG-0448. The sensitivity as a function of two parameters, $S(q,s)$, is shown on the left panel, and on the right is shown the integrated sensitivity $S(q)$ when a flat distribution of $s$ in $\log{s}$ is assumed. In both panels we show the sensitivities of two solutions $(-+)$ (solid) and $(--)$ (dashed). 
\label{fig:sensitivity}}
\end{figure}

\begin{figure}
\includegraphics[width=.81\textwidth]{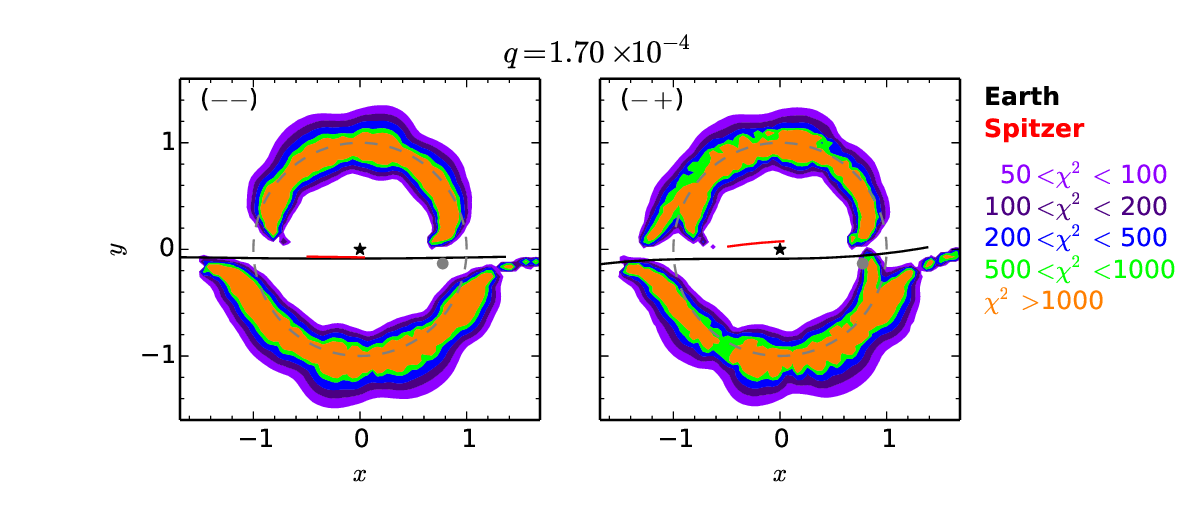}
\caption{The $\chi^2$ distributions of simulated OGLE-2015-BLG-0448 light curves with a $q=1.7\times10^{-4}$ planet placed at different positions $(x,y)$. The left panel shows the result for the $(--)$ solution, and the right panel shows that for the $(-+)$ solution. The black/red lines indicate the source trajectories as seen from Earth/\emph{Spitzer}. 
The lens is placed at $(0,0)$, and the position of the tentative planet is shown as a filled gray dot. 
Note that the tentative planet could only be detected in the $(-+)$ solution.
\label{fig:chisq-map}}
\end{figure}

\begin{figure}
\includegraphics[width=.81\textwidth,bb= 19 480 576 700]{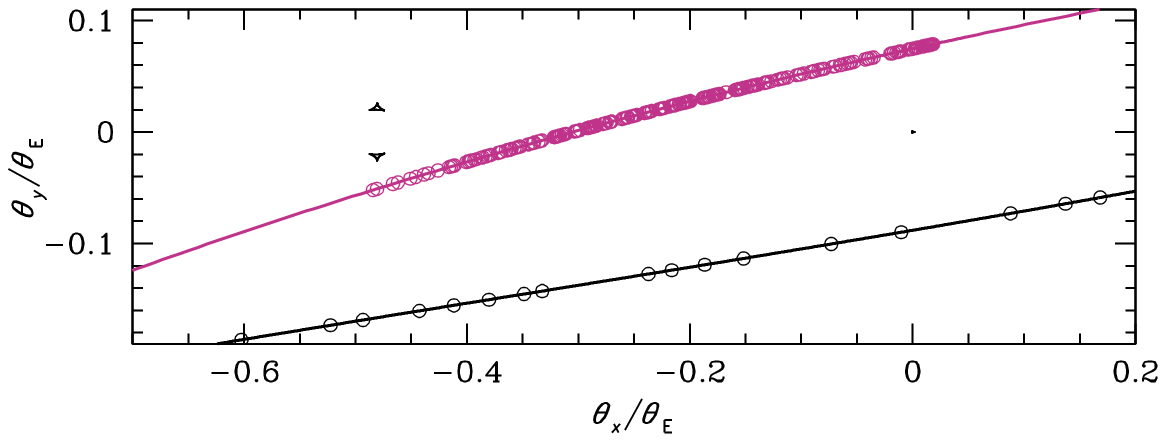}
\caption{Source trajectory as seen from \emph{Spitzer} (violet) and Earth (black). The central caustic is located at $(\theta_x,\theta_y) = (0,0)$ and two triangular planetary caustics are at $\theta_x/\theta_\e = s - 1/s \approx -0.48$. The circles indicate apparent source positions at the epoch when \emph{Spitzer} and OGLE data were taken. 
\label{fig:traj}}
\end{figure}

\end{document}